# Enabling and Inhibitory Pathways of Students' AI Use Concealment Intention in Higher Education: Evidence from SEM and fsQCA

**First Author and Corresponding Author**
Yiran Du
University of Cambridge, Cambridge, UK
yd392@cam.ac.uk

**Second Author**
Huimin He
Xi'an Jiaotong-Liverpool University, Suzhou, China
Huimin.he@xjtlu.edu.cn

**Abstract**
This study investigates students' AI use concealment intention in higher education by integrating the cognition–affect–conation (CAC) framework with a dual-method approach combining structural equation modelling (SEM) and fuzzy-set qualitative comparative analysis (fsQCA). Drawing on data from 1,346 university students, the findings reveal two opposing mechanisms shaping concealment intention. The enabling pathway shows that perceived stigma, perceived risk, and perceived policy uncertainty increase fear of negative evaluation, which in turn promotes concealment. In contrast, the inhibitory pathway demonstrates that AI self-efficacy, perceived fairness, and perceived social support enhance psychological safety, thereby reducing concealment intention. SEM results confirm the hypothesised relationships and mediation effects, while fsQCA identifies multiple configurational pathways, highlighting equifinality and the central role of fear of negative evaluation across conditions. The study contributes to the literature by conceptualising concealment as a distinct behavioural outcome and by providing a nuanced explanation that integrates both net-effect and configurational perspectives. Practical implications emphasise the need for clear institutional policies, destigmatisation of appropriate AI use, and the cultivation of supportive learning environments to promote transparency.

## 1. Introduction

Artificial intelligence (AI) is rapidly reconfiguring higher education by changing how students learn, produce academic work, and respond to assessment demands. Generative AI tools can enhance efficiency, autonomy, and personalised learning, while educators are increasingly adopting AI for adaptive instruction, automated feedback, and learning analytics (Qian, 2025; C. Du et al., 2025; Y. Du et al., 2026). At the same time, the growing use of AI has generated concerns about academic integrity, ethical boundaries, and unclear institutional rules, creating an environment in which students may perceive AI use as both academically useful and socially risky (Wiese et al., 2025; Yan et al., 2025).

Within this context, an emerging but insufficiently examined issue is students' intention to conceal their use of AI in academic tasks. AI use concealment intention refers to students' deliberate intention to hide or obscure their reliance on AI tools, often to avoid negative judgement, reputational harm, or academic sanctions (BaHammam, 2025; Stone, 2025). Existing research has primarily focused on AI adoption, acceptance, and continuance use, but has paid less attention to concealment as a distinct behavioural outcome. More importantly, limited research has explained how affective mechanisms shape this intention. In particular, fear of negative evaluation may increase concealment intention because students who anticipate unfavourable judgement are more likely to withhold disclosure, whereas psychological safety may reduce concealment intention because students who feel safe in their learning environment are less likely to hide their AI use (Ans et al., 2026; Bao & Zeng, 2026; X. Zhang et al., 2025).

To address this gap, the present study draws on the cognition–affect–conation framework to explain how cognitive appraisals generate affective responses that subsequently influence concealment intention (Zhou & Zhang, 2024). Specifically, this study proposes a dual-path model. In the enabling pathway, perceived stigma, perceived risk, and perceived policy uncertainty are expected to heighten fear of negative evaluation, which in turn increases AI use concealment intention. In the inhibitory pathway, AI self-efficacy, perceived fairness, and perceived social support are expected to strengthen psychological safety, which in turn decreases AI use concealment intention. By combining structural equation modelling with fuzzy-set qualitative comparative analysis, this study extends current scholarship by providing both net-effect and configurational explanations of AI use concealment intention in higher education.

## 2. Literature Review
### 2.1 AI in Higher Education
Artificial intelligence (AI) has rapidly become a transformative force in higher education, reshaping teaching, learning, and assessment practices (Qian, 2025). Tools such as generative AI systems enable students to access instant explanations, generate academic content, and personalise learning pathways, thereby enhancing efficiency and autonomy (C. Du et al., 2025). At the same time, educators are increasingly integrating AI into instructional design, using it to support adaptive learning, automate feedback, and analyse student performance data (Y. Du et al., 2026). However, the growing presence of AI also raises critical concerns related to academic integrity, ethical use, and policy ambiguity, as institutions struggle to establish clear guidelines for appropriate use (Wiese et al., 2025). These tensions position AI as a double-edged technology: while it offers significant pedagogical and productivity benefits, it simultaneously introduces risks associated with overreliance, misuse, and unequal access (Adarkwah et al., 2025). Consequently, understanding how students perceive and engage with AI in higher education has become an important area of inquiry, particularly in relation to emerging behaviours such as the concealment of AI use.

### 2.2 AI Use Concealment Intention
AI use concealment intention in this study refers to students' deliberate efforts to hide or obscure their use of artificial intelligence tools in academic tasks. This behavioural intention is often conceptualised as a self-protective response shaped by evaluative concerns, where individuals anticipate negative judgement from instructors or peers and thus choose non-disclosure (BaHammam, 2025). Drawing on emerging literature, concealment intention is influenced by cognitive appraisals (e.g., perceived risk, policy uncertainty, and stigma), affective responses (e.g., fear of negative evaluation), and contextual factors such as classroom norms and institutional policies (Ans et al., 2026; Bao & Zeng, 2026). In higher education settings, where expectations around originality and authorship remain salient, students may experience tension between the instrumental benefits of AI use and the potential social or academic consequences of disclosure (X. Zhang et al., 2025). As a result, concealment becomes a strategic behaviour aimed at mitigating reputational or academic risk (Stone, 2025). Understanding AI use concealment intention is therefore critical for unpacking how students navigate the evolving ethical landscape of AI-assisted learning and for informing the development of clearer, more supportive governance frameworks.

## 3. Theoretical Framework and Hypothesis Development
### 3.1 Cognition–Affect–Conation Framework
The cognition–affect–conation (CAC) framework provides a useful theoretical lens for explaining how individuals' cognitive appraisals shape affective responses, which in turn influence behavioural intention (Zhou & Zhang, 2024). In the present study, cognition is reflected in students' perceptions of AI use in higher education, including perceived stigma, perceived risk, perceived policy uncertainty, AI self-efficacy, perceived fairness, and perceived social support; affect is reflected in fear of negative evaluation and psychological safety; and conation is represented by AI use concealment intention. The framework is particularly appropriate for this research because it could capture both enabling and inhibitory mechanisms through which cognitive appraisals may either intensify or reduce concealment intention via distinct affective states. Based on this logic, the conceptual model is proposed in Figure 1, while the justification for the selected factors and the corresponding hypotheses is presented in Sections 3.2 and 3.3.

**Figure 1. The Conceptual Model**

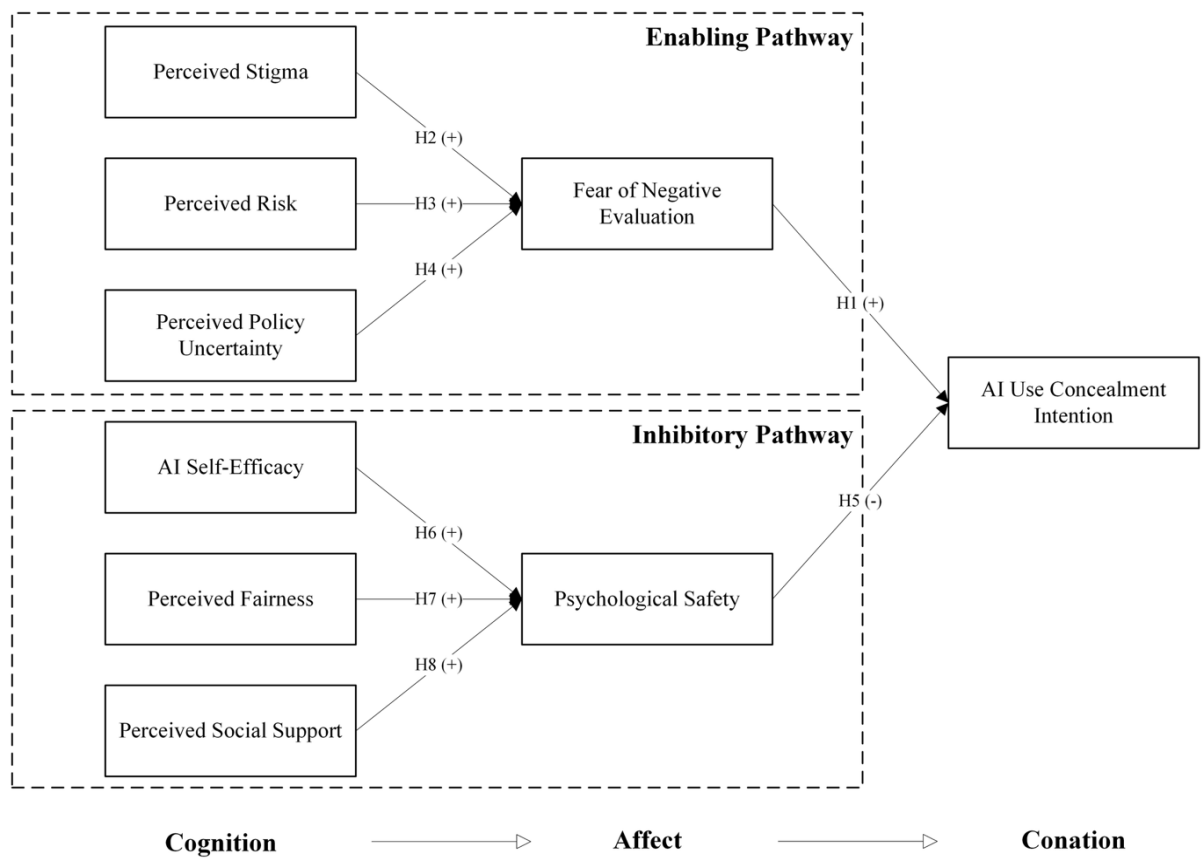

## 3.2 The Enabling Pathway of AI Use Concealment Intention

Fear of negative evaluation is conceptualised as the central affective mechanism in the enabling pathway, referring to individuals' apprehension about being judged unfavourably by others (Yue et al., 2022). Grounded in social evaluation theory and extensively validated in educational psychology, this construct reflects individuals' sensitivity to external judgement in contexts where norms and performance expectations are salient (Jia & Yue, 2023). In higher education, the increasing yet contested use of AI tools introduces ambiguity around acceptable practices, which may heighten students' concern about being perceived as dishonest, incompetent, or overly reliant on technology (Yan et al., 2025). Prior research has consistently shown that fear of negative evaluation is a key driver of impression management and self-protective behaviours, including avoidance, non-disclosure, and concealment (Gao et al., 2025; Nor et al., 2025; Y. Zhang et al., 2025). In this study, it is therefore positioned as the immediate antecedent of AI use concealment intention, capturing how affective concern is translated into behavioural intention.

Three cognitive antecedents are proposed to explain the emergence of fear of negative evaluation: perceived stigma, perceived risk, and perceived policy uncertainty. These constructs are selected because they represent three distinct but complementary dimensions of threat appraisal associated with AI use, social evaluation, anticipated consequences, and institutional ambiguity (Van Den Bos & Lind, 2002). Perceived stigma in this study refers to the extent to which students believe that using AI is socially disapproved of or morally questionable; according to stigma theory, such perceptions increase individuals' concern about being negatively judged and socially devalued (Hatzenbuehler et al., 2024; Thornicroft et al., 2022). Perceived risk in this study captures students' assessment of potential negative outcomes, such as academic penalties, loss of credibility, or damage to reputation; risk perception theory suggests that anticipated adverse consequences heighten anxiety and evaluative concern (Gaube et al., 2019; Scherer & Cho, 2003). Perceived policy uncertainty in this study reflects ambiguity or inconsistency in institutional guidelines governing AI use; drawing on uncertainty reduction and organisational behaviour literature, unclear rules are likely to intensify evaluative anxiety as individuals lack reliable standards to guide their behaviour (S. Li et al., 2025; Milliken, 1987). Empirically, prior studies on academic misconduct and emerging technologies indicate that stigma, risk, and policy ambiguity are salient predictors of evaluative concern and defensive behavioural responses (Newton, 2025; Nguyen & Goto, 2024; Smit et al., 2025; Stone, 2025). Taken together, these cognitive appraisals are expected to increase fear of negative evaluation, which in turn promotes AI use concealment intention. Accordingly, the following hypotheses are proposed:

H1: Fear of negative evaluation is positively associated with AI use concealment intention.
H2: Perceived stigma is positively associated with fear of negative evaluation.
H3: Perceived risk is positively associated with fear of negative evaluation.
H4: Perceived policy uncertainty is positively associated with fear of negative evaluation.

### 3.3 The Inhibitory Pathway of AI Use Concealment Intention
The inhibitory pathway explains how supportive cognitive appraisals of AI use in higher education give rise to protective affective states that reduce students' intention to conceal their use of AI. Within the cognition–affect–conation framework, psychological safety is conceptualised as the central affective mechanism, referring to individuals' perception that they can engage in a behaviour without fear of negative consequences to their self-image, status, or relationships (Edmondson & Bransby, 2023). Originating from organisational behaviour research, psychological safety has been widely applied in educational contexts to explain openness, participation, and willingness to take interpersonal risks (Dong et al., 2025; Lackie et al., 2023). In the context of AI use, when students perceive their learning environment as supportive and non-punitive, they are less likely to experience evaluative anxiety and more likely to disclose or openly engage with AI tools (Ans et al., 2026). Accordingly, psychological safety is expected to function as a key inhibitor of AI use concealment intention.

Three cognitive antecedents are proposed to explain the development of psychological safety: AI self-efficacy, perceived fairness, and perceived social support. These constructs are selected because they capture complementary resource-based appraisals that foster confidence, trust, and a sense of security in using AI. AI self-efficacy refers to students' belief in their capability to effectively and appropriately use AI tools (Wang & Chuang, 2024); grounded in self-efficacy theory, stronger perceptions of competence are expected to reduce uncertainty and increase confidence in one's actions, thereby fostering a greater sense of psychological safety (Waddington, 2023). Perceived fairness in this study reflects the extent to which institutional policies and instructional practices regarding AI use are viewed as just, transparent, and consistently applied; fairness theory suggests that such perceptions enhance trust and reduce defensiveness, thereby enhancing psychological safety (Lind, 2019; Nicklin et al., 2011). Perceived social support in this study denotes the extent to which students feel supported by instructors and peers in their use of AI; social support theory indicates that supportive environments buffer stress and promote psychological safety (Alshammari & Alkhwaldi, 2025; Vaux, 1988). Empirical research in educational and organisational settings further demonstrates that competence beliefs, fair treatment, and social support are key predictors of psychological safety and open behaviour (Jindal et al., 2024; Pinho & Colston, 2025). Taken together, these cognitive appraisals are expected to enhance psychological safety, which in turn reduces AI use concealment intention. Accordingly, the following hypotheses are proposed:

H5: Psychological safety is negatively associated with AI use concealment intention.
H6: AI self-efficacy is positively associated with psychological safety.
H7: Perceived fairness is positively associated with psychological safety.
H8: Perceived social support is positively associated with psychological safety.

### 4. Methods
### 4.1 Research Design
The study adopted a quantitative, cross-sectional research design to examine the enabling and inhibitory mechanisms underlying AI use concealment intention among university students. Grounded in the cognition–affect–conation framework, the research employed a survey-based approach to collect self-reported data on multiple latent constructs, enabling the empirical testing of hypothesised relationships within a structured conceptual model. To capture both linear and configurational causal patterns, a mixed-analytical strategy was implemented, combining structural equation modelling (SEM) to assess net effects and mediation pathways, with fuzzy-set qualitative comparative analysis (fsQCA) to identify equifinal configurations of conditions leading to concealment intention. This design allows for a comprehensive examination of both symmetric and asymmetric relationships, thereby providing a more nuanced understanding of the phenomenon.

### 4.2 Participants

Participants were recruited through Credamo, a widely used Chinese online survey platform. A total of 1,489 responses were initially collected from Chinese university students. After screening for careless responding, including failed attention-check items, excessively short completion times, and invariant response patterns, 143 responses were excluded (Ward & Meade, 2023). The final sample therefore comprised 1,346 Chinese university students. As shown in Appendix A, the sample was relatively balanced in terms of gender and academic discipline. In addition, most participants were aged between 18 and 23 years, and the majority were undergraduate students. Ethical approval was obtained from the author's institutional ethics committee prior to data collection. All participants provided informed consent before taking part and were free to withdraw at any time without penalty.

### 4.3 Measurement

All constructs, perceived stigma (Warren, 2023), perceived risk (C. Zhang et al., 2025), perceived policy uncertainty (S. Li et al., 2025), fear of negative evaluation (Bozkurt, 2025), AI self-efficacy (Wang & Chuang, 2024), perceived fairness (Chambers et al., 2023), perceived social support (D. Li et al., 2025), psychological safety (McGuire, 2025), and AI use concealment intention (Ozuna & Steinhoff, 2024), were measured using established multi-item scales adapted from prior literature and tailored to the context of AI use in higher education (Y. Du, 2024), All items were rated on a five-point Likert scale ranging from 1 (strongly disagree) to 5 (strongly agree). The items were originally developed in English and translated into Chinese following a back-translation procedure to ensure semantic equivalence (Brislin, 1970). Minor wording adjustments were made following a pilot study ($N = 50$). The full list of measurement items is presented in Appendix B.

Several procedural and statistical measures were adopted to minimise common method variance (CMV) (Podsakoff et al., 2024). Participants were assured of anonymity and confidentiality to reduce evaluation apprehension and social desirability bias. The questionnaire used clear wording, and the focal constructs were organised into separate sections to reduce the likelihood that participants would infer expected relationships among variables. In addition, some items were reverse-coded and varied scale formats were used where appropriate to limit response pattern bias. Harman's single-factor test showed that the first unrotated factor accounted for 12.3% of the total variance, suggesting that common method variance was unlikely to be a serious concern.

### 4.4 Data Analysis

Data were analysed using a combination of structural equation modelling (SEM) (Kline, 2023) and fuzzy-set qualitative comparative analysis (fsQCA) (Schneider & Wagemann, 2012). SEM was conducted using a two-step approach to assess the measurement model and the structural model. First, confirmatory factor analysis (CFA) was performed to evaluate reliability and validity, including indicator loadings, Cronbach's α, composite reliability (CR), and average variance extracted (AVE), as well as discriminant validity using the Fornell–Larcker criterion and the heterotrait–monotrait (HTMT) ratio. Model fit was assessed using multiple indices, including $\chi^2/df$, CFI, TLI, RMSEA, and SRMR. Second, the structural model was examined to test the hypothesised relationships, with standardised path coefficients and significance levels estimated. Mediation effects were assessed using bootstrapping with 5,000 resamples and bias-corrected confidence intervals. To complement the net-effect perspective of SEM, fsQCA was employed to examine configurational pathways leading to AI use concealment intention. Following established procedures, construct scores were calibrated into fuzzy sets using percentile-based anchors, and necessity and sufficiency analyses were conducted using consistency and coverage thresholds.

### 5. Results

## 5.1 Structural Equation Modelling (SEM)

Descriptive statistics for all constructs are reported in Appendix C. Overall, the results indicate that participants reported relatively higher levels of perceived risk, fear of negative evaluation, and concealment intention, whereas enabling factors such as perceived fairness, social support, and psychological safety were comparatively lower. The standard deviations suggest moderate variability across all constructs. Skewness and kurtosis values fell within acceptable ranges (skewness: −0.58 to 0.68; kurtosis: −0.63 to 0.42), indicating no serious deviation from normality.

The measurement model was evaluated in terms of reliability, validity, and overall model fit. As reported in Appendix E, all indicator loadings exceeded the recommended threshold of 0.70, indicating satisfactory indicator reliability. Cronbach's α and composite reliability (CR) values for all constructs were above 0.70, demonstrating adequate internal consistency. In addition, the average variance extracted (AVE) values ranged from 0.63 to 0.73, exceeding the recommended threshold of 0.50 and supporting convergent validity. Discriminant validity was established using both the Fornell–Larcker criterion and the HTMT ratio. As shown in Appendix F, the square roots of the AVE were greater than the inter-construct correlations, and all HTMT values in Appendix G were below the conservative threshold of 0.85. Furthermore, the measurement model exhibited a good fit to the data ($\chi^2/df = 2.18$, CFI = 0.94, TLI = 0.93, RMSEA = 0.046, SRMR = 0.041), with all indices meeting recommended threshold values. Together, these results indicate that the measurement model demonstrates satisfactory reliability and validity with an acceptable model fit.

The structural model results are reported in Figure 2 and Appendix H, and the model fit indices are presented in Appendix D. Overall, the model demonstrated an acceptable fit to the data ($\chi^2/df = 2.36$, CFI = 0.92, TLI = 0.91, RMSEA = 0.051, SRMR = 0.048). The model also explained a meaningful proportion of variance in the endogenous constructs, with $R^2 = 0.49$ for fear of negative evaluation, $R^2 = 0.41$ for psychological safety, and $R^2 = 0.46$ for AI use concealment intention. Fear of negative evaluation was positively associated with AI use concealment intention ($\beta = 0.38$, $p < 0.001$), supporting H1. Perceived stigma ($\beta = 0.29$, $p < 0.001$), perceived risk ($\beta = 0.34$, $p < 0.001$), and perceived policy uncertainty ($\beta = 0.18$, $p < 0.001$) were all positively associated with fear of negative evaluation, supporting H2, H3, and H4. In addition, psychological safety was negatively associated with concealment intention ($\beta = -0.22$, $p < 0.001$), supporting H5. AI self-efficacy ($\beta = 0.28$, $p < 0.001$), perceived fairness ($\beta = 0.19$, $p < 0.01$), and perceived social support ($\beta = 0.24$, $p < 0.001$) were positively associated with psychological safety, supporting H6, H7, and H8. These findings support both the enabling and inhibitory pathways proposed in the conceptual model.

The results of the mediation analysis are presented in Appendix I. Using bootstrapping procedures (5,000 resamples), the indirect effects were found to be statistically significant, as all 95% confidence intervals excluded zero. Specifically, fear of negative evaluation significantly mediated the effects of perceived stigma (indirect effect = 0.11, 95% CI [0.07, 0.16]), perceived risk (indirect effect = 0.13, 95% CI [0.09, 0.18]), and perceived policy uncertainty (indirect effect = 0.07, 95% CI [0.03, 0.11]) on AI use concealment intention. In addition, psychological safety significantly mediated the relationships between AI self-efficacy (indirect effect = −0.06, 95% CI [−0.10, −0.02]), perceived fairness (indirect effect = −0.04, 95% CI [−0.08, −0.01]), and perceived social support (indirect effect = −0.05, 95% CI [−0.09, −0.02]) and concealment intention. These findings provide further support for the underlying mechanisms proposed in the conceptual model.

**Figure 2. Structural Model Results**

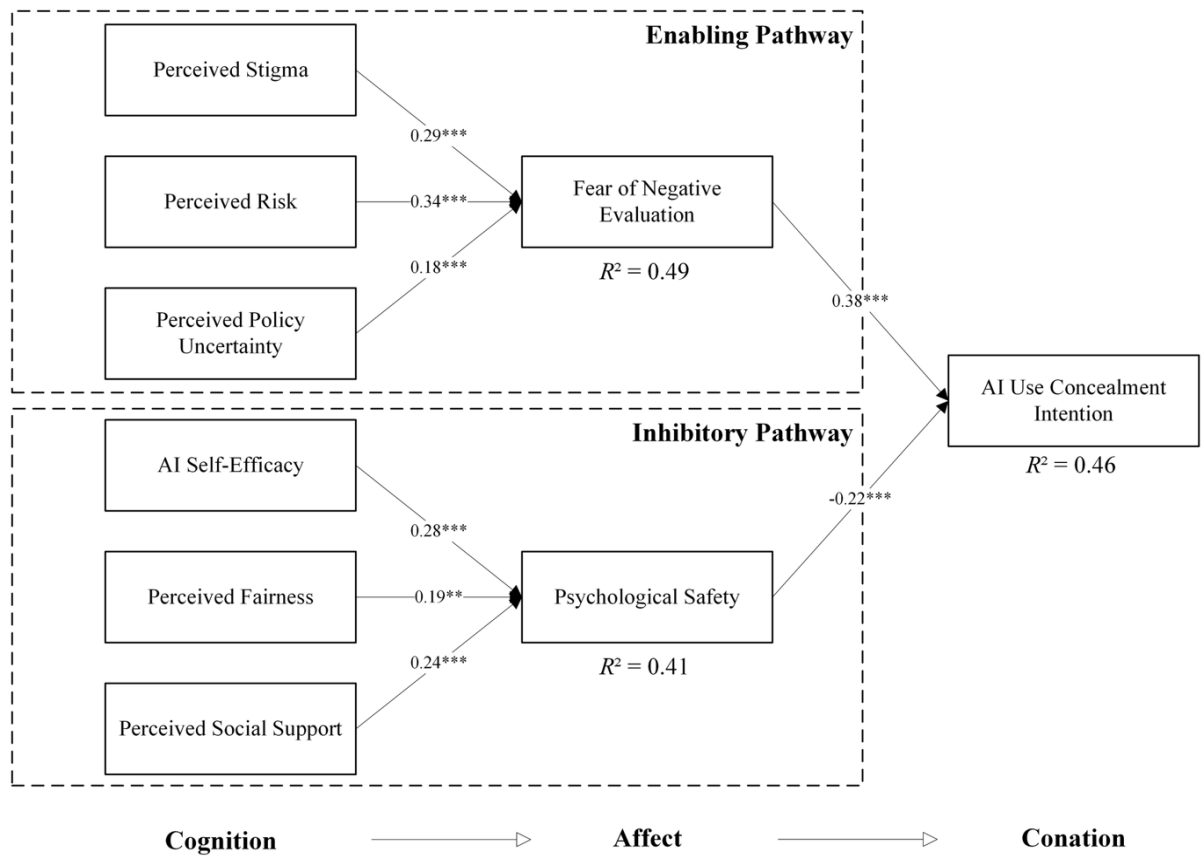

Note. Standardised path coefficients are reported. *** $p < 0.001$; ** $p < 0.01$. $R^2$ values indicate the proportion of variance explained in the endogenous constructs.

### 5.2 Fuzzy-Set Qualitative Comparative Analysis (fsQCA)
Following the SEM analysis, fsQCA was employed to complement the net-effect approach by examining configurational causality. Prior to the analysis, all construct scores were aggregated and calibrated into fuzzy sets using three qualitative anchors: the 95th percentile for full membership, the 50th percentile for the crossover point, and the 5th percentile for full non-membership. A necessity analysis was conducted, and none of the antecedent conditions reached the recommended consistency threshold of 0.90, indicating that no single condition constituted a necessary condition for AI use concealment intention. A truth table was then constructed, with a frequency threshold of 8, a consistency threshold of 0.80, and a proportional reduction in inconsistency (PRI) threshold of 0.65.

The fsQCA results are presented in Table 1. Three configurational pathways leading to high AI use concealment intention were identified, with an overall solution consistency of 0.81 and coverage of 0.52, indicating acceptable explanatory power. Across all configurations, fear of negative evaluation emerged as a core or peripheral condition, underscoring its central role in driving concealment intention. Two configurations represent threat-based mechanisms characterised by different combinations of perceived stigma, perceived risk, and perceived policy uncertainty, together with fear of negative evaluation and reduced psychological safety. A third configuration reflects a resource-deficit pathway, in which low perceived fairness, low perceived social support, low AI self-efficacy, and low psychological safety, together with fear of negative evaluation, contribute to concealment intention. These findings further reveal equifinality, suggesting that AI use concealment intention can arise from multiple combinations of conditions rather than a single linear causal path.

**Table 1. Configuration Paths of AI Use Concealment Intention**

| Conditions | Path 1 | Path 2 | Path 3 |
|---|---|---|---|
| Perceived Stigma | ○ | ● | |
| Perceived Risk | ● | ○ | |
| Perceived Policy Uncertainty | ○ | ● | |
| Fear of Negative Evaluation | ● | ● | ○ |
| AI Self-Efficacy | | | ⊖ |
| Perceived Fairness | | | ⊗ |
| Perceived Social Support | | | ⊗ |
| Psychological Safety | ⊗ | ○ | ⊗ |
| Consistency | 0.84 | 0.82 | 0.83 |
| Raw coverage | 0.29 | 0.24 | 0.20 |
| Unique coverage | 0.08 | 0.05 | 0.04 |
| Overall solution consistency | | 0.81 | |
| Overall solution coverage | | 0.52 | |

Note. ● indicates the presence of a core condition; ○ indicates the presence of a peripheral condition; ⊗ indicates the absence of a core condition; ⊖ indicates the absence of a peripheral condition; blank cells indicate "don't care" conditions.

## 6. Discussion
### 6.1 The Enabling Pathway of AI Use Concealment Intention
The findings of this study provide empirical support for the enabling pathway proposed within the cognition–affect–conation framework, demonstrating that fear of negative evaluation serves as a critical affective mechanism linking threat-related cognitive appraisals to AI use concealment intention. Specifically, perceived stigma, perceived risk, and perceived policy uncertainty were all found to significantly increase fear of negative evaluation, which in turn strongly predicted concealment intention. This pattern is consistent with prior research suggesting that individuals who anticipate negative judgement or social devaluation are more likely to engage in self-protective behaviours such as non-disclosure and concealment (Gao et al., 2025; Nor et al., 2025; Y. Zhang et al., 2025). In the context of AI use in higher education, where norms remain ambiguous and ethical concerns are salient, students may interpret AI use as potentially illegitimate or risky, thereby heightening their sensitivity to external evaluation (Yan et al., 2025; Stone, 2025). The strong effect of perceived risk further aligns with risk perception theory, which posits that anticipated negative consequences amplify evaluative anxiety and defensive behavioural responses (Gaube et al., 2019; Scherer & Cho, 2003). Similarly, perceived stigma reinforces concerns about social judgement, as students who believe AI use is morally questionable are more likely to fear reputational damage (Hatzenbuehler et al., 2024; Thornicroft et al., 2022).

In addition, the role of perceived policy uncertainty highlights the importance of institutional ambiguity in shaping affective responses and behavioural intentions. When students lack clear guidance on acceptable AI use, uncertainty increases their vulnerability to evaluative concerns, as they are unable to rely on stable norms to justify their behaviour (Milliken, 1987; S. Li et al., 2025). This finding is consistent with emerging research indicating that unclear or inconsistent AI policies may contribute to moral ambiguity and defensive strategies, including concealment (Smit et al., 2025; Newton, 2025). The mediating role of fear of negative evaluation therefore confirms its function as a proximal psychological driver that translates diverse cognitive threat appraisals into concealment intention. Overall, the enabling pathway underscores that AI use concealment is not merely a rational response to external constraints but is deeply rooted in affective processes associated with social evaluation and uncertainty. These results extend prior literature on academic integrity and technology use by demonstrating how multiple forms of perceived threat converge through a shared emotional mechanism to shape students' behavioural intentions in AI-mediated learning environments

### 6.2 The Inhibitory Pathway of AI Use Concealment Intention
The findings provide robust support for the inhibitory pathway, demonstrating that psychological safety functions as a key affective mechanism that reduces AI use concealment intention. Consistent with the cognition–affect–conation framework, the results indicate that AI self-efficacy, perceived fairness, and perceived social support positively contribute to psychological safety, which in turn significantly decreases concealment intention. This aligns with prior research suggesting that individuals who perceive their environment as supportive and non-threatening are more willing to engage in open behaviours and less likely to adopt defensive strategies such as concealment (Edmondson & Bransby, 2023; Dong et al., 2025). In the context of AI use in higher education, when students feel confident in their ability to use AI appropriately and perceive institutional practices as fair and transparent, they are less likely to experience evaluative anxiety and more likely to disclose their AI use. This finding also resonates with emerging studies on AI disclosure, which indicate that supportive environments reduce the psychological barriers associated with transparency (Ans et al., 2026; Bao & Zeng, 2026).

Furthermore, the positive effects of AI self-efficacy, perceived fairness, and perceived social support highlight the importance of resource-based cognitive appraisals in fostering psychological safety. Drawing on self-efficacy theory, students with stronger beliefs in their AI-related competence are better equipped to navigate uncertainty and are less prone to defensive behaviour (Wang & Chuang, 2024; Waddington, 2023). Similarly, fairness perceptions enhance trust in institutional systems and reduce perceptions of arbitrariness, thereby promoting a sense of security (Lind, 2019; Nicklin et al., 2011). Social support from instructors and peers further reinforces this effect by buffering stress and normalising AI use within the learning environment (Alshammari & Alkhwaldi, 2025; Vaux, 1988). Together, these findings suggest that reducing AI use concealment intention requires not only mitigating perceived threats but also actively cultivating supportive, fair, and competence-enhancing environments. By empirically validating psychological safety as a central inhibitor, this study extends prior literature on educational technology and academic integrity, demonstrating that openness in AI use is facilitated by affective assurance grounded in positive cognitive appraisals

## 6.3 Configurational Mechanisms of AI Use Concealment Intention

The fsQCA results provide important complementary insights by revealing the configurational mechanisms underlying AI use concealment intention, highlighting that this behaviour is characterised by equifinality rather than a single linear causal path. Specifically, three distinct configurations were identified, each representing different combinations of cognitive and affective conditions that lead to high concealment intention. Across these configurations, fear of negative evaluation consistently emerged as a central or peripheral condition, reinforcing its pivotal role identified in the SEM results. This finding aligns with prior research suggesting that evaluative concern is a core driver of self-protective and impression management behaviours across contexts (Gao et al., 2025; Nor et al., 2025). However, the configurational analysis extends this understanding by demonstrating that fear of negative evaluation does not operate in isolation but interacts with multiple contextual and cognitive factors to produce concealment intention. This supports the broader configurational perspective that behavioural outcomes are often the result of interdependent conditions rather than independent net effects (Schneider & Wagemann, 2012).

More specifically, two configurations reflect threat-dominant pathways, where different combinations of perceived stigma, perceived risk, and perceived policy uncertainty jointly contribute to concealment intention through heightened fear of negative evaluation and reduced psychological safety. These findings reinforce theoretical arguments that multiple forms of perceived threat, social, institutional, and outcome-related, can substitute for one another in generating similar behavioural responses (Van Den Bos & Lind, 2002; Smit et al., 2025). In contrast, the third configuration represents a resource-deficit pathway, characterised by low AI self-efficacy, low perceived fairness, low social support, and low psychological safety. This pathway suggests that concealment intention can also arise in the absence of strong threat perceptions, instead emerging from a lack of supportive resources and confidence. Such a pattern is consistent with research emphasising the role of competence beliefs, fairness, and social support in fostering openness and reducing defensive behaviour (Wang & Chuang, 2024; Alshammari & Alkhwaldi, 2025). Taken together, these findings demonstrate that AI use concealment intention is a multifaceted phenomenon shaped by both threat-based and resource-based configurations, thereby extending existing literature by integrating symmetric and asymmetric explanations within a unified analytical framework

## 6.4 Theoretical and Practical Implications

The present study offers several important theoretical contributions to the literature on AI use in higher education and academic integrity. First, it extends the cognition–affect–conation (CAC) framework by explicitly modelling both enabling and inhibitory pathways of AI use concealment intention, thereby providing a more balanced and process-oriented explanation of student behaviour. While prior research has largely focused on AI adoption and continuance use (Qian, 2025; C. Du et al., 2025), this study shifts attention to concealment as a distinct and theoretically meaningful outcome. By demonstrating that fear of negative evaluation and psychological safety function as dual affective mechanisms, the findings advance understanding of how opposing emotional processes mediate the relationship between cognitive appraisals and behavioural intention (Zhou & Zhang, 2024). Second, the study contributes to emerging research on AI ethics and disclosure by identifying stigma, risk, and policy uncertainty as key antecedents of evaluative concern, thereby enriching theoretical accounts of why students may avoid transparency in AI use (Yan et al., 2025; Bao & Zeng, 2026). Third, by integrating SEM with fsQCA, the study responds to calls for methodological pluralism and demonstrates that AI-related behaviours are shaped by both net effects and configurational causality, offering a more nuanced and holistic explanation of concealment intention (Schneider & Wagemann, 2012).

From a practical perspective, the findings suggest that reducing AI use concealment intention requires coordinated interventions targeting both threat perceptions and supportive conditions. Educational institutions should prioritise the development of clear, consistent, and transparent AI policies to reduce policy uncertainty and minimise students' fear of negative evaluation (Smit et al., 2025; Newton, 2025). At the same time, efforts should be made to destigmatise appropriate AI use through curriculum design, assessment reform, and open dialogue, thereby addressing the social and moral concerns associated with AI-assisted work (Wiese et al., 2025). Importantly, the results also highlight the need to foster psychologically safe learning environments by enhancing AI self-efficacy, ensuring fairness in evaluation practices, and strengthening social support from instructors and peers. Such interventions can reduce defensive behaviours and encourage responsible disclosure of AI use (Edmondson & Bransby, 2023; Alshammari & Alkhwaldi, 2025). Overall, the study provides actionable insights for educators and policymakers seeking to balance the benefits of AI integration with the need to uphold academic integrity, emphasising that both structural clarity and affective assurance are essential for promoting transparent and ethical AI use in higher education

## 6.5 Limitations and Future Directions

Several limitations should be acknowledged when interpreting the findings of this study. First, the use of a cross-sectional design limits the ability to establish causal relationships among cognitive appraisals, affective mechanisms, and AI use concealment intention. Although the proposed model is theoretically grounded, the temporal ordering of these variables cannot be definitively confirmed. Second, the reliance on self-reported data introduces the possibility of response biases, particularly social desirability bias, given that concealment behaviours involve sensitive and potentially undesirable actions. Despite efforts to mitigate common method bias, the use of a single data source may still inflate observed relationships. Third, the sample was restricted to university students within a single national context, which may limit the generalisability of the findings. Cultural, institutional, and policy differences across regions may influence how students perceive AI use and whether they choose to disclose or conceal it.

Future research can address these limitations by employing longitudinal or experimental designs to better establish causal mechanisms and examine how AI use concealment intention evolves over time. Incorporating multi-source or behavioural data would also strengthen the robustness of findings and reduce reliance on self-reports. In addition, cross-cultural and cross-institutional studies are needed to assess the generalisability of the proposed model and to identify context-specific dynamics. Future work could further extend the model by integrating additional psychological and contextual variables, such as moral judgement, academic motivation, or disciplinary norms, which may shape students' decisions regarding AI use. Qualitative approaches, including interviews or case studies, may also provide deeper insights into the nuanced reasoning processes underlying concealment behaviour, thereby complementing the quantitative findings of this study.

## 7. Conclusion

This study advances understanding of AI use concealment intention in higher education by integrating the cognition–affect–conation framework with both symmetric (SEM) and configurational (fsQCA) analyses. The findings demonstrate that concealment intention is shaped by dual mechanisms: an enabling pathway in which perceived stigma, risk, and policy uncertainty heighten fear of negative evaluation, and an inhibitory pathway in which AI self-efficacy, perceived fairness, and social support enhance psychological safety and reduce concealment. By evidencing both net effects and equifinal configurations, the study shows that concealment is not driven by a single factor but emerges from multiple interacting conditions, with fear of negative evaluation playing a central role across pathways. These results contribute theoretically by positioning concealment as a distinct behavioural outcome in AI research and by highlighting the importance of affective mechanisms, while also offering practical implications for fostering transparent AI use through clearer policies, reduced stigma, and more supportive learning environments.

**Appendices**

**Appendix A. Participant Characteristics (*N* = 1346)**

| Characteristic | Category | n | % |
|---|---|---|---|
| Gender | Male | 622 | 46.2 |
|  | Female | 724 | 53.8 |
| Age | 18–20 | 489 | 36.3 |
|  | 21–23 | 634 | 47.1 |
|  | 24 or above | 223 | 16.6 |
| Study level | Undergraduate | 1038 | 77.1 |
|  | Postgraduate | 308 | 22.9 |
| Academic discipline | STEM | 676 | 50.2 |
|  | Non-STEM | 670 | 49.8 |

**Appendix B Constructs and Measurement Items**

| Construct | Item | Measurement Item (English) | Measurement Item (Chinese) |
|---|---|---|---|
| Perceived Stigma (Warren, 2023) | PS1 | People would think less of me if I used AI tools in my coursework. | 如果我在课程作业中使用 AI 工具，别人会看低我。 |
|  | PS2 | Using AI tools for academic work is negatively viewed by others. | 在学术任务中使用 AI 工具会受到他人的负面看待。 |
|  | PS3 | Students who use AI tools in academic work are viewed unfavourably. | 在学术任务中使用 AI 工具的学生通常会被他人负面看待。 |
| Perceived Risk (Zhang et al., 2025) | PR1 | Using AI tools could lead to academic penalties. | 使用 AI 工具可能会带来学术惩罚。 |
|  | PR2 | Using AI tools in coursework could negatively affect my academic outcomes. | 在课程作业中使用 AI 工具可能会对我的学业结果产生不利影响。 |
|  | PR3 | Using AI tools in academic work is risky. | 在学术任务中使用 AI 工具是有风险的。 |
| Perceived Policy Uncertainty (S. Li et al., 2025) | PPU1 | I am unclear about my university's rules on AI use. | 我不清楚学校关于 AI 使用的规定。 |
|  | PPU2 | My university's policies on AI use are ambiguous. | 学校关于 AI 使用的政策是模糊的。 |
|  | PPU3 | It is difficult to determine what kinds of AI use are allowed. | 我很难判断哪些类型的 AI 使用是被允许的。 |
| Fear of Negative Evaluation (Bozkurt, 2025) | FNE1 | I worry that others would evaluate me negatively if I used AI tools. | 如果我使用 AI 工具，我担心他人会对我作出负面评价。 |
|  | FNE2 | I feel anxious that others may disapprove of my use of AI tools. | 我会担心他人可能不认可我使用 AI 工具。 |
|  | FNE3 | I am concerned about being criticised for using AI tools. | 我担心因使用 AI 工具而受到批评。 |

| Construct | Item | English | Chinese |
|---|---|---|---|
| AI Self-Efficacy (Wang & Chuang, 2024) | ASE1 | I am confident in my ability to use AI tools effectively. | 我有信心能够有效地使用 AI 工具。 |
| | ASE2 | I can use AI tools effectively for my academic tasks. | 我能够有效地将 AI 工具用于学术任务。 |
| | ASE3 | I can learn to use new AI tools when needed. | 在需要时，我能够学会使用新的 AI 工具。 |
| Perceived Fairness (Chambers et al., 2023) | PF1 | My university's rules on AI use are fair. | 我认为学校关于 AI 使用的规定是公平的。 |
| | PF2 | The policies on AI use are applied consistently across students. | 关于 AI 使用的政策在不同学生之间的执行是一致的。 |
| | PF3 | My university's AI policies do not unfairly disadvantage students. | 学校的 AI 政策不会对学生造成不公平的不利影响。 |
| Perceived Social Support (D. Li et al., 2025) | PSS1 | People around me support my use of AI tools for learning. | 我周围的人支持我在学习中使用 AI 工具。 |
| | PSS2 | I feel supported by others when I use AI tools in my studies. | 当我在学习中使用 AI 工具时，我能感受到他人的支持。 |
| | PSS3 | Others approve of my use of AI tools for academic purposes. | 他人认可我将 AI 工具用于学术任务。 |
| Psychological Safety (McGuire, 2025) | PSY1 | I feel safe using AI tools in my academic work. | 我在学业中使用 AI 工具时感到安全。 |
| | PSY2 | I do not feel threatened when using AI tools in my studies. | 我在学习中使用 AI 工具时不会感到受威胁。 |
| | PSY3 | I feel comfortable trying AI tools in my studies. | 我在学习中尝试使用 AI 工具时感到自在。 |
| AI Use Concealment Intention (Ozuna & Steinhoff, 2024) | AUCI1 | I intend to conceal my use of AI tools in coursework. | 我打算隐瞒自己在课程作业中使用 AI 工具的行为。 |
| | AUCI2 | I plan not to disclose my use of AI tools to instructors. | 我打算不向教师透露自己使用 AI 工具的情况。 |
| | AUCI3 | I intend to keep my use of AI tools private. | 我打算对自己使用 AI 工具的情况保密。 |

**Appendix C. Descriptive Statistics of the Constructs**

| Construct | $M$ | SD | Skewness | Kurtosis |
|---|---|---|---|---|
| Perceived Stigma | 3.62 | 0.84 | -0.42 | 0.18 |
| Perceived Risk | 3.81 | 0.79 | -0.58 | 0.42 |
| Perceived Policy Uncertainty | 3.55 | 0.87 | -0.21 | -0.63 |
| Fear of Negative Evaluation | 3.69 | 0.83 | -0.47 | 0.25 |
| AI Self-Efficacy | 3.12 | 0.88 | 0.36 | -0.48 |
| Perceived Fairness | 2.98 | 0.91 | 0.52 | -0.15 |
| Perceived Social Support | 3.05 | 0.89 | 0.41 | -0.37 |
| Psychological Safety | 2.91 | 0.94 | 0.68 | 0.12 |
| AI Use Concealment Intention | 3.67 | 0.86 | -0.50 | 0.31 |

**Appendix D. Model Fit Indices**

| Fit Index | Threshold | Measurement Model | Structural Model |
|---|---|---|---|
| $\chi^2$/df | < 3.00 | 2.18 | 2.36 |
| CFI | > 0.90 | 0.94 | 0.92 |
| TLI | > 0.90 | 0.93 | 0.91 |
| RMSEA | < 0.08 | 0.046 | 0.051 |
| SRMR | < 0.08 | 0.041 | 0.048 |

Note. χ²/df = chi-square divided by degrees of freedom; CFI = Comparative Fit Index; TLI = Tucker-Lewis Index; RMSEA = Root Mean Square Error of Approximation; SRMR = Standardised Root Mean Square Residual.

**Appendix E. Reliability and Convergent Validity**

| Construct | Item | Loading | Cronbach's α | CR | AVE |
|---|---|---|---|---|---|
| Perceived Stigma | PS1 | 0.81 | 0.84 | 0.85 | 0.65 |
| | PS2 | 0.79 | | | |
| | PS3 | 0.82 | | | |
| Perceived Risk | PR1 | 0.84 | 0.86 | 0.87 | 0.69 |
| | PR2 | 0.81 | | | |
| | PR3 | 0.85 | | | |
| Perceived Policy Uncertainty | PPU1 | 0.80 | 0.83 | 0.84 | 0.63 |
| | PPU2 | 0.77 | | | |
| | PPU3 | 0.81 | | | |
| Fear of Negative Evaluation | FNE1 | 0.83 | 0.85 | 0.86 | 0.67 |
| | FNE2 | 0.79 | | | |
| | FNE3 | 0.83 | | | |
| AI Self-Efficacy | ASE1 | 0.85 | 0.87 | 0.88 | 0.71 |
| | ASE2 | 0.83 | | | |
| | ASE3 | 0.85 | | | |
| Perceived Fairness | PF1 | 0.82 | 0.84 | 0.85 | 0.66 |
| | PF2 | 0.79 | | | |
| | PF3 | 0.82 | | | |
| Perceived Social Support | PSS1 | 0.81 | 0.83 | 0.84 | 0.63 |
| | PSS2 | 0.78 | | | |
| | PSS3 | 0.80 | | | |
| Psychological Safety | PSY1 | 0.84 | 0.86 | 0.87 | 0.69 |
| | PSY2 | 0.81 | | | |
| | PSY3 | 0.85 | | | |
| AI Use Concealment Intention | AUCI1 | 0.86 | 0.88 | 0.89 | 0.73 |
| | AUCI2 | 0.84 | | | |
| | AUCI3 | 0.86 | | | |

Note. CR = composite reliability; AVE = average variance extracted.

**Appendix F. Discriminant Validity (Fornell–Larcker Criterion)**

| Construct | PS | PR | PPU | FNE | ASE | PF | PSS | PSY | AUCI |
|---|---|---|---|---|---|---|---|---|---|
| PS | **0.81** | | | | | | | | |
| PR | 0.48 | **0.83** | | | | | | | |
| PPU | 0.42 | 0.46 | **0.79** | | | | | | |
| FNE | 0.55 | 0.58 | 0.49 | **0.82** | | | | | |
| ASE | -0.26 | -0.29 | -0.22 | -0.31 | **0.84** | | | | |
| PF | -0.31 | -0.34 | -0.27 | -0.35 | 0.41 | **0.81** | | | |
| PSS | -0.28 | -0.30 | -0.25 | -0.32 | 0.44 | 0.46 | **0.79** | | |
| PSY | -0.33 | -0.36 | -0.29 | -0.38 | 0.48 | 0.50 | 0.53 | **0.83** | |
| AUCI | 0.52 | 0.57 | 0.48 | 0.60 | -0.37 | -0.42 | -0.40 | -0.45 | **0.85** |

Note. Diagonal elements in bold represent the square root of the average variance extracted (AVE). PS = Perceived Stigma; PR = Perceived Risk; PPU = Perceived Policy Uncertainty; FNE = Fear of Negative Evaluation; ASE = AI Self-Efficacy; PF = Perceived Fairness; PSS = Perceived Social Support; PSY = Psychological Safety; AUCI = AI Use Concealment Intention.

**Appendix G. Discriminant Validity (HTMT Ratio)**

| Construct | PS | PR | PPU | FNE | ASE | PF | PSS | PSY | AUCI |
|---|---|---|---|---|---|---|---|---|---|

|     | PS | PR | PPU | FNE | ASE | PF | PSS | PSY | AUCI |
|-----|----|----|-----|-----|-----|----|-----|-----|------|
| PS  | —  |    |     |     |     |    |     |     |      |
| PR  | 0.62 | — |     |     |     |    |     |     |      |
| PPU | 0.55 | 0.58 | — |   |     |    |     |     |      |
| FNE | 0.71 | 0.74 | 0.63 | — |  |    |     |     |      |
| ASE | 0.33 | 0.36 | 0.29 | 0.38 | — | |    |     |      |
| PF  | 0.39 | 0.42 | 0.34 | 0.44 | 0.52 | — |  |     |      |
| PSS | 0.36 | 0.38 | 0.32 | 0.41 | 0.56 | 0.58 | — |  |      |
| PSY | 0.41 | 0.45 | 0.36 | 0.48 | 0.61 | 0.64 | 0.67 | — |   |
| AUCI | 0.69 | 0.73 | 0.60 | 0.76 | 0.45 | 0.50 | 0.48 | 0.54 | — |

**Note.** HTMT = heterotrait–monotrait ratio of correlations. PS = Perceived Stigma; PR = Perceived Risk; PPU = Perceived Policy Uncertainty; FNE = Fear of Negative Evaluation; ASE = AI Self-Efficacy; PF = Perceived Fairness; PSS = Perceived Social Support; PSY = Psychological Safety; AUCI = AI Use Concealment Intention.

**Appendix H. Structural Model Results**

| Hypothesis | Path | β | SE | z | Result |
|------------|------|---|----|----|--------|
| H1 | FNE → AUCI | 0.38 | 0.05 | 7.60*** | Supported |
| H2 | PS → FNE | 0.29 | 0.04 | 7.25*** | Supported |
| H3 | PR → FNE | 0.34 | 0.05 | 7.01*** | Supported |
| H4 | PPU → FNE | 0.18 | 0.05 | 3.60*** | Supported |
| H5 | PSY → AUCI | -0.22 | 0.06 | -3.67*** | Supported |
| H6 | ASE → PSY | 0.28 | 0.05 | 5.60*** | Supported |
| H7 | PF → PSY | 0.19 | 0.06 | 3.17** | Supported |
| H8 | PSS → PSY | 0.24 | 0.06 | 4.00*** | Supported |

Note. β = standardised path coefficient; SE = standard error; z = critical ratio. *** $p < 0.001$; ** $p < 0.01$. PS = Perceived Stigma; PR = Perceived Risk; PPU = Perceived Policy Uncertainty; FNE = Fear of Negative Evaluation; ASE = AI Self-Efficacy; PF = Perceived Fairness; PSS = Perceived Social Support; PSY = Psychological Safety; AUCI = AI Use Concealment Intention.

**Appendix I. Mediation Analysis Results (Bootstrapping)**

| Path | Indirect Effect | Boot SE | 95% CI (LL, UL) | Result |
|------|-----------------|---------|-----------------|--------|
| PS → FNE → AUCI | 0.11 | 0.02 | [0.07, 0.16] | Significant |
| PR → FNE → AUCI | 0.13 | 0.02 | [0.09, 0.18] | Significant |
| PPU → FNE → AUCI | 0.07 | 0.02 | [0.03, 0.11] | Significant |
| ASE → PSY → AUCI | -0.06 | 0.02 | [-0.10, -0.02] | Significant |
| PF → PSY → AUCI | -0.04 | 0.02 | [-0.08, -0.01] | Significant |
| PSS → PSY → AUCI | -0.05 | 0.02 | [-0.09, -0.02] | Significant |

Note. Bootstrapping based on 5,000 resamples. CI = confidence interval; LL = lower limit; UL = upper limit. An indirect effect is significant if the confidence interval does not include zero.